\documentclass[12pt]{revtex4-1}
\usepackage{amsmath}
\usepackage{amssymb}
\usepackage[left=1cm,right=1cm]{geometry}
\usepackage[colorlinks,linkcolor=blue,citecolor=red]{hyperref}
\usepackage{feynmp}
\usepackage{slashed}
\usepackage{color}
\usepackage{graphicx}
\usepackage{makecell}
\usepackage{braket}

\begin{document}

\title{\LARGE Path Integral Method for Proportional Step and Proportional Double-Barrier Step Option Pricing}
\bigskip
\author{Qi Chen~$^{1}$}
\email{20080142038@lfnu.edu.cn}
\author{Chao Guo~$^{2}$}
\email{chaog@lfnu.edu.cn}
\affiliation{
$^{1}$~School of Economics and Management, Langfang Normal University, Langfang 065000, China
\\
$^2$ School of Science, Langfang Normal University, Langfang 065000, China
}
\date{\today}

\begin{abstract}
Path integral method in quantum mechanics provides a new thinking for barrier option pricing. For proportional step options, the option price changing process is similar to the one dimensional trapezoid potential barrier scattering problem in quantum mechanics; for double-barrier step options, the option price changing process is analogous to a particle moving in a finite symmetric square potential well. Using path integral method, the analytical expressions of pricing kernel and option price could be derived. Numerical results of option price as a function of underlying price, potential and exercise price are shown, which are consistent with the results given by mathematical method. 
\end{abstract}

\maketitle

\section{Introduction}
In 1973, Black and Scholes derived the analytical expression for fixed-volatility option price by solving stochastic differential equations ~\cite{Black}. Financial mathematics applied to to derivative pricing has made great progress from then on ~\cite{Amin,Rubinstein}. Recently, more complex options have emerged in financial market, which collectively called exotic options ~\cite{Merton:1973,Geske:1979}. A kind of exotic option is the one attached some conditions to an ordinary option, taking barrier options for example: when the underlying price touches this barrier, the option contract will be activated, which is called knock-in option; when the underlying price touches the barrier, the option contract is invalid, which is call knock-out option. Snyder had discussed down-and-out option in 1969 ~\cite{Snyder}. Baaquie et al discussed this kind of option by path integral method, and derived the corresponding analytical expressions ~\cite{Baaquie}. Similar to one-dimensional infinite square potential well in quantum mechanics, the analytical expression for double-knock-out option price was also derived~\cite{Baaquie:2004}, which is in accordance with the result derived by mathematical method ~\cite{Haug}. In addition, path integral method has been applied to the research of interest rate derivative pricing ~\cite{Marakani, Srikant}. Early works investigating step options appear in~\cite{Linetsky:2001, Linetsky:1999}. Studies of option pricing by path integral method can be found in~\cite{Baaquie, Baaquie:2004, Kleinert:2012}.

In this paper, we will discuss two kinds of barrier options, which are called proportional step option and proportional double-barrier step option, respectively. A step option could be also called a gradual knock-out (knock-in) option: when the underlying price touches and passes the barrier, the option contract is not invalid immediately, but the option knocks out (knocks in) gradually ~\cite{Linetsky:2001,Linetsky:1999}. 
A proportional step call is defined by its payoff~\cite{Linetsky:1999}
 \begin{equation}
     e^{-V_0\tau} {\rm max}(S_T-K,0)
 \end{equation}
 where $S_T$ is the underlying price at expiration date, $K$ is the exercise price, $V_0$ is called knock-out rate, which is corresponding to the potential in quantum mechanics. The exponential $e^{-V_0\tau}$ is interpreted as a knock-out discount factor with discounting time $\tau$. A daily knock-out factor $\beta$ could be defined as
 \begin{equation}
     \beta=e^{-V_0/250}
 \end{equation}
 where 250 trading days per year is assumed. We focus on the relation between proportional step call (proportional double-barrier step call) option pricing and one dimensional trapezoid potential scattering (finite symmetric square well): when a particle moving ahead and going through the boundary, the wave function begins to decay exponentially, which is similar to a proportional step (proportional double-barrier step) option knocks out over time. 

Our work is organized as follows. In Section 2, we will derive the analytical expressions for proportional step call option price by path integral method. Approximate analytical expressions for proportional double-barrier step call price are given in Section 3, In Section 4,  we show the numerical results for option price as a function of underlying price, exercise price and potential, respectively. We summarize our main results in Section 5. The pricing formulas for Black-Scholes model, the standard barrier option and the standard double-barrier option derived by path integral method are reviewed in Appendix A, B and C~\cite{Baaquie:2004}.

\section{Proportional Step Option Pricing}
  An up-and-out proportional step call price changing could be analogous to a one-dimensional particle moving in the following potential
\begin{equation}\label{potential1}
V(x)=\left\{
\begin{aligned}
0 & , & x<B,\\
V_0 & , & x\geq B.
\end{aligned}
\right.
\end{equation}
for $x> B$, the wave function decays as the distance increases in the case of the particle energy $E<V_0$, which is similar to an option touches a barrier at $x=B$ and knocks out gradually. Making the following variable substitution~\cite{Baaquie}
\begin{equation}
    S=e^x,\ x\in (-\infty, +\infty)
\end{equation}
the Hamiltonian for a proportional step option (PSO) is
\begin{equation}\label{Hamil}
    H_{\rm PSO}=-\frac{\sigma^2}{2}\frac{\partial^2}{\partial x^2}+\left(\frac{1}{2}\sigma^2-r\right)\frac{\partial}{\partial x}+r+V(x)
\end{equation}
where $S\in(0,+\infty)$ is the underlying price, $r$ is the fixed interest rate, and $\sigma$ is the volatility. The Hamiltonian (\ref{Hamil}) is non-Hermitian, considering the following transformation
\begin{equation}
    H_{\rm PSO}=e^{\alpha x}H_{\rm eff}e^{-\alpha x}=e^{\alpha x}\left(-\frac{\sigma^2}{2}\frac{\partial^2}{\partial x^2}+\gamma\right)e^{-\alpha x}+V(x) 
\end{equation}
where
\begin{equation}
    \alpha=\frac{1}{\sigma^2}\left(\frac{\sigma^2}{2}-r\right),\ \ \gamma=\frac{1}{2\sigma^2}\left(\frac{\sigma^2}{2}+r\right)^2
\end{equation}
and $H_{\rm eff}$ is Hermitian. The stationary state Schr{\"o}dinger equation for option price is 
\begin{equation}\label{schrodinger}
    \left\{
\begin{aligned}
-\frac{\sigma^2}{2}\frac{{\rm d}^2 C}{{\rm d}x^2}+\gamma C=EC & , & x<B,\\
-\frac{\sigma^2}{2}\frac{{\rm d}^2 C}{{\rm d}x^2}+(\gamma+V_0) C=EC & , & x\geq B.
\end{aligned}
\right.
\end{equation}
where $C$ is the option price, and $E$ is corresponding to the particle energy under the potential (\ref{potential1}). The Schr{\"o}dinger equation above could be simplified into 
\begin{equation}
    \left\{
\begin{aligned}\label{simpleschrodinger1}
\frac{{\rm d}^2 C}{{\rm d}x^2}+p_1^2 C=0 & , & x<B,\\
\frac{{\rm d}^2 C}{{\rm d}x^2}-p_2^2 C=0 & , & x\geq B.
\end{aligned}
\right.
\end{equation}
where
\begin{equation}
    p_1^2=\frac{2(E-\gamma)}{\sigma^2},\ \ p_2^2=\frac{2(V_0+\gamma-E)}{\sigma^2}
\end{equation}
the condition $E<V_0$ has been considered, and the range of $p_1^2$ and $p_2^2$ is 
\begin{equation}\begin{split}
    p_2^2&>\frac{2\gamma}{\sigma^2}\\
    p_1^2&=\frac{2V_0}{\sigma^2}-p_2^2<\frac{2(V_0-\gamma)}{\sigma^2}
\end{split}\end{equation}

The general solution for (\ref{simpleschrodinger1}) is
\begin{equation}\label{solution}
    C(x)=
    \left\{
\begin{aligned}
&e^{ip_1(x-B)}+ A_1e^{-ip_1(x-B)} , & x<B,\\
&A_2e^{-p_2(x-B)}  , & x\geq B.
\end{aligned}
\right.
\end{equation}
where
\begin{equation}
    A_1=\frac{p_1-ip_2}{p_1+ip_2},\ \ A_2=\frac{2p_1}{p_1+ip_2}
\end{equation}
here the boundary condition at $x=B$ is used. Now we calculate the price of up-and-out proportional step call. Let $\tau_1$ indicates the occupation time below the barrier $B$, and $\tau_2$ is the occupation time above the barrier $B$. The pricing kernel is 
\begin{equation}\begin{split}
    p_{\rm PSO}(x,x^\prime;\tau)&=\braket{x|e^{-\tau_1H_1-\tau_2H_2}|x^\prime}\\
    &=e^{\alpha(x-x^\prime)}\int\frac{{\rm d}p}{2\pi}\braket{x|e^{-\tau_1 H_{\rm eff1}}|p}\braket{p|e^{-\tau_2 H_{\rm eff_2}}|x^\prime}
\end{split}\end{equation}
and the option price is
\begin{equation}\label{optionprice}
    C(x;\tau)=\int_{{\rm ln}K}^{+\infty}{\rm d}x^\prime p_{\rm PSO}(x,x^\prime;\tau)(e^{x^\prime}-K)
\end{equation}
where
\begin{equation}\begin{split}
    H_1&=-\frac{\sigma^2}{2}\frac{\partial^2}{\partial x^2}+\left(\frac{1}{2}\sigma^2-r\right)\frac{\partial}{\partial x}+r\\
    H_2&=-\frac{\sigma^2}{2}\frac{\partial^2}{\partial x^2}+\left(\frac{1}{2}\sigma^2-r\right)\frac{\partial}{\partial x}+r+V_0\\
    H_{\rm eff1}&=-\frac{\sigma^2}{2}\frac{\partial^2}{\partial x^2}+\gamma\\
    H_{\rm eff2}&=-\frac{\sigma^2}{2}\frac{\partial^2}{\partial x^2}+\gamma+V_0
\end{split}\end{equation}

We set the exercise price $K<e^B$. For $x<B$ and $x>B$, the wave functions are different, and the integral for (\ref{optionprice}) should be split into four cases
\begin{itemize}
	\item $x<B$, and\ \  ${\rm ln}K<x^\prime<B$
\end{itemize}
the pricing kernel
\begin{equation}\begin{split}
   p_{{\rm PSO1}}(x,x^\prime;\tau)&=e^{-\tau\gamma}e^{\alpha(x-x^\prime)}\int_0^{\frac{\sqrt{2(V_0-\gamma)}}{\sigma}}\frac{{\rm d}p_1}{2\pi}e^{-\frac{1}{2}\tau\sigma^2p_1^2}\bigg(e^{ip_1(x-B)}+A_1e^{-ip_1(x-B)}\bigg)\times\\
   &\bigg(e^{-ip_1(x^\prime-B)}+A_1^{*}e^{ip_1(x^\prime-B)}\bigg)\\
   &=e^{-\tau\gamma}e^{\alpha(x-x^\prime)}\int_0^{\frac{\sqrt{2(V_0-\gamma)}}{\sigma}}\frac{{\rm d}p_1}{2\pi}e^{-\frac{1}{2}\tau\sigma^2p_1^2}\bigg[2\cos p_1(x-x^\prime)+\frac{\sigma^2}{V_0}(p_1^2-p_2^2)\cos p_1(x+x^\prime-2B)-\\
   &\frac{2\sigma^2}{V_0}p_1p_2\sin p_1(x+x^\prime-2B)\bigg]
\end{split}\end{equation}
\begin{itemize}
	\item $x<B$, and\ \  $x^\prime>B$
\end{itemize}
the pricing kernel
\begin{equation}\begin{split}
   p_{{\rm PSO2}}(x,x^\prime;\tau)&=e^{-\tau\gamma}e^{\alpha(x-x^\prime)}\int_0^{\frac{\sqrt{2(V_0-\gamma)}}{\sigma}}\frac{{\rm d}p_1}{2\pi}e^{-\frac{1}{2}\tau\sigma^2p_1^2}\bigg(e^{ip_1(x-B)}+A_1e^{-ip_1(x-B)}\bigg)A_2^{*}e^{-p_2(x^\prime-B)}\\
&=2e^{-\tau\gamma}e^{\alpha(x-x^\prime)}\int_0^{\frac{\sqrt{2(V_0-\gamma)}}{\sigma}}\frac{{\rm d}p_1}{2\pi}e^{-\frac{1}{2}\tau\sigma^2p_1^2}\bigg[\frac{\sigma^2}{V_0}p_1^2\cos p_1(x-B)-\frac{\sigma^2}{V_0}p_1p_2\sin p_1(x-B)\bigg]\times\\
&e^{-p_2(x^\prime-B)}
\end{split}\end{equation}
the option price for $x<B$ is 
\begin{equation}
    C(x;\tau)|_{x<B}=C_1(x;\tau)+C_2(x;\tau)
\end{equation}
where
\begin{equation}\begin{split}
     C_1(x,\tau)&=\int_{{\rm ln}K}^B p_{{\rm PSO1}}(x,x^\prime;\tau)(e^{x^\prime}-K)\\
     C_2(x,\tau)&=\int_{B}^{+\infty} p_{{\rm PSO2}}(x,x^\prime;\tau)(e^{x^\prime}-K)
\end{split}\end{equation}
\begin{itemize}
	\item $x>B$, and\ \  ${\rm ln}K<x^\prime<B$
\end{itemize}
the pricing kernel
\begin{equation}\begin{split}
    p_{{\rm PSO3}}(x,x^\prime;\tau)&=e^{-\tau\gamma}e^{\alpha(x-x^\prime)}\int_0^{\frac{\sqrt{2(V_0-\gamma)}}{\sigma}}\frac{{\rm d}p_1}{2\pi}e^{-\frac{1}{2}\tau\sigma^2p_1^2}
A_2e^{-p_2(x-B)}\bigg(e^{-ip_1(x^\prime-B)}+A_1^{*}e^{ip_1(x^\prime-B)}\bigg)\\
&=2e^{-\tau\gamma}e^{\alpha(x-x^\prime)}\int_0^{\frac{\sqrt{2(V_0-\gamma)}}{\sigma}}\frac{{\rm d}p_1}{2\pi}e^{-\frac{1}{2}\tau\sigma^2p_1^2}e^{-p_2(x-B)}\bigg[\frac{\sigma^2}{V_0}p_1^2\cos p_1(x^\prime-B)\\
&-\frac{\sigma^2}{V_0}p_1p_2\sin p_1(x^\prime-B)\bigg]
\end{split}\end{equation}
\begin{itemize}
	\item $x>B$, and\ \  $x^\prime>B$
\end{itemize}
\begin{equation}\begin{split}
    p_{{\rm PSO4}}(x,x^\prime;\tau)&=e^{-\tau\gamma}e^{\alpha(x-x^\prime)}\int_0^{\frac{\sqrt{2(V_0-\gamma)}}{\sigma}}\frac{{\rm d}p_1}{2\pi}e^{-\frac{1}{2}\tau\sigma^2p_1^2}
A_2e^{-p_2(x-B)}A_2^{*}e^{-p_2(x^\prime-B)}\\
&=2e^{-\tau\gamma}e^{\alpha(x-x^\prime)}\int_0^{\frac{\sqrt{2(V_0-\gamma)}}{\sigma}}\frac{{\rm d}p_1}{2\pi}e^{-\frac{1}{2}\tau\sigma^2p_1^2}\frac{\sigma^2}{V_0}p_1^2e^{-p_2(x+x^\prime-2B)}
\end{split}\end{equation}
the option price for $x>B$ is
\begin{equation}
    C(x;\tau)|_{x>B}=C_3(x;\tau)+C_4(x;\tau)
\end{equation}
where
\begin{equation}\begin{split}
   C_3(x,\tau)&=\int_{{\rm ln}K}^B p_{{\rm PSO3}}(x,x^\prime;\tau)(e^{x^\prime}-K)\\
   C_4(x,\tau)&=\int_{B}^{+\infty} p_{{\rm PSO4}}(x,x^\prime;\tau)(e^{x^\prime}-K)
\end{split}\end{equation}
\section{Proportional Double-Barrier Step Option Pricing}

The price changing of a proportional double-barrier step ({\rm PDBS}) option could be analogous to a particle moving in a symmetric square potential well with the potential
\begin{equation}\label{potential}
V(x)=\left\{
\begin{aligned}
0 & , & a<x<b,\\
V_0 & , & x<a,\ x>b.
\end{aligned}
\right.
\end{equation}
for $x<a$ or $x>b$, the wave function decays with the increasing distances from the well, which is similar to an option touches a barrier and knocks out gradually. The Hamitonian for a double-barrier step option is~\cite{Baaquie}
\begin{equation}\label{nonhermitehami}
H_{\rm PDBS}=-\frac{\sigma^2}{2}\frac{\partial^2}{\partial x^2}+\left(\frac{1}{2}\sigma^2-r\right)\frac{\partial}{\partial x}+r+V(x)
\end{equation}
which is a non-Hermitian Hamitonian. Considering the following transformation
\begin{equation}\label{hermitehami}
H_{\rm PDBS}=e^{\alpha x} H_{\rm {eff}}e^{-\alpha x}=
e^{\alpha x}\left(-\frac{\sigma^2}{2}\frac{\partial^2}{\partial x^2}+\gamma\right)e^{-\alpha x}+V(x)
\end{equation}
where
\begin{equation}
\alpha=\frac{1}{\sigma^2}\left(\frac{\sigma^2}{2}-r\right),\\ 
\gamma=\frac{1}{2\sigma^2}\left(\frac{\sigma^2}{2}+r\right)^2
\end{equation}
and $H_{\rm {eff}}$ is a Hermitian Hamitonian which is considered as the symmetric square potential well Hamitonian. The stationary state Schr{\"o}dinger equation for option price is
\begin{equation}
    \left\{
\begin{aligned}\label{schrodingereq}
&-\frac{\sigma^2}{2}\frac{{\rm d}^2 C}{{\rm d}x^2}+\gamma C=EC  , & a<x<b,\\
&-\frac{\sigma^2}{2}\frac{{\rm d}^2 C}{{\rm d}x^2}+(\gamma+V_0) C=EC  , & x<a,\ x>b.
\end{aligned}
\right.
\end{equation}
where $C$ is the option price, $E$ is corresponding to bound state energy levels in the potential well. (\ref{schrodingereq}) could be simplified into
\begin{equation}
    \left\{
\begin{aligned}\label{simpleschrodinger}
\frac{{\rm d}^2 C}{{\rm d}x^2}+k_1^2 C=0 & , & a<x<b,\\
\frac{{\rm d}^2 C}{{\rm d}x^2}-k_2^2 C=0 & , & x<a,\ x>b.
\end{aligned}
\right.
\end{equation}
where 
\begin{equation}
    k_1^2=\frac{2(E-\gamma)}{\sigma^2},\ \ k_2^2=\frac{2(V_0+\gamma-E)}{\sigma^2}
\end{equation}

The general solution for (\ref{simpleschrodinger}) is
\begin{equation}\label{general solution}
C(x)=\left\{
\begin{aligned}
A_3\ e^{k_2(x-\frac{b+a}{2})} & , & x\leq a, \\
A_1 \sin(k_1x+\delta) & , & a< x \leq b, \\
A_2\ e^{-k_2(x-\frac{b+a}{2})} & , & x>b.
\end{aligned}
\right.
\end{equation}
Now we will use the method provided in~\cite{HZY:2018} to derive the approximate energy level formulas for (\ref{general solution}). Considering the continuity for both wave function and its derivative at $x=a$ and $x=b$, we have 
\begin{equation}\label{delta value}
    \delta=\frac{\ell\pi}{2}-k_1\frac{b+a}{2},\ \ \ell=0, 1, 2,...
\end{equation}

According to different $\ell s$ in (\ref{delta value}), (\ref{general solution}) could be split into two parts

\begin{equation}\label{oddeq}
C_1(x)=\left\{
\begin{aligned}
&-A_2e^{k_2(x-\frac{b+a}{2})}  , & x\leq a, \\
&A_1 \sin k_1\left(x-\frac{b+a}{2}\right) , & a\leq x \leq b,\\ 
&A_2 e^{-k_2(x-\frac{b+a}{2})}  , & x>b.
\end{aligned}
\right.\ \ \ \ for\ \ \ell=0, 2, 4,...
\end{equation}
and
\begin{equation}\label{eveneq}
C_2(x)=\left\{
\begin{aligned}
&A_2 e^{k_2(x-\frac{b+a}{2})}  , & x\leq a, \\
&A_1 \cos k_1\left(x-\frac{b+a}{2}\right)  , & a\leq x \leq b, \\
&A_2 e^{-k_2(x-\frac{b+a}{2})}  , & x>b.
\end{aligned}
\right.\ \ \ \ for\ \ \ell=1, 3, 5,...
\end{equation}	
where 
\begin{equation}\label{A1A2}
    A_1=\sqrt{\frac{2k_2}{k_2(b-a)+2}},\ \ A_2=A_1\sin k_1\frac{b-a}{2}e^{k_2\frac{b-a}{2}}
\end{equation}
here the normalization condition has been used. Considering boundary conditions for (\ref{oddeq}) and (\ref{eveneq}) at $x=b$ respectively, we have
\begin{equation}\label{condition1}
    \cot k_1\frac{b-a}{2}=-\frac{k_2}{k_1}
    \end{equation}
    \begin{equation}\label{condition2}
    \tan k_1\frac{b-a}{2}=\frac{k_2}{k_1}
\end{equation}
let 
\begin{equation}\label{beta}
    \theta=\arcsin\frac{k_1}{\beta}\in\left(0,\ \frac{\pi}{2}\right),\ \beta=\sqrt{k_1^2+k_2^2}=\frac{\sqrt{2V_0}}{\sigma}
\end{equation}
(\ref{condition1}) and (\ref{condition2}) could be combined into
\begin{equation}\label{conditionk1}
    k_{1n}\frac{b-a}{2}=\frac{n\pi}{2}-\theta,\ \ n=1,2,3,...
\end{equation}
allowing for $\theta\in(0,\pi/2)$, the range of $k_{1n}$ is
\begin{equation}
    \frac{(n-1)\pi}{b-a}<k_{1n}<\frac{n\pi}{b-a}
\end{equation}
when $n\to n_{max}$, the energy $E_n\approx V_0$, and
\begin{equation}
    k_{1n}=\frac{2(E_n-\gamma)}{\sigma}\to \sqrt{\beta^2-\frac{2\gamma}{\sigma^2}}\approx \frac{n_{max}\pi}{b-a}
\end{equation}
where $n_{max}$ is the maximum number of energy levels, and
\begin{equation}\label{maximumn}
    n_{max}=\left[\frac{b-a}{\pi}\sqrt{\beta^2-\frac{2\gamma}{\sigma^2}}\right]
\end{equation}
$[x]$ indicates the minimal integer not less than $x$. In general, there is no analytical solution for energy eigenvalues. For low energy case $(E\ll V_0)$, considering only the first order approximation of (\ref{conditionk1}), 
\begin{equation}
    k_{1n}\frac{b-a}{2}=\frac{n\pi}{2}-\arcsin{\frac{k_{1n}}{\beta}}\approx\frac{n\pi}{2}-\frac{k_{1n}}{\beta} 
\end{equation}
and the low energy level formula is
\begin{equation}\label{loweq}
    k_{1n}\approx\frac{\beta n\pi}{\beta(b-a)+2}
\end{equation}
the error of $k_{1n}$ is
\begin{equation}
\Delta k_{1n}\approx \frac{1}{6}\left(\frac{k_{1n}}{\beta}\right)^3=\frac{n^3\pi^3}{6[\beta(b-a)+2]^3}
\end{equation}
where $\mathcal{O} (k_{1n}^5)$ and higher orders have been ignored. The relative error for $k_{1n}$ is
\begin{equation}
    \delta k_{1n}=\bigg|\frac{\Delta k_{1n}}{k_{1n}}\bigg|=\frac{n^2\pi^2}{6\beta[\beta(b-a)+2]^2}
\end{equation}

For high energy case $(E_n\approx V_0)$, the approximation of (\ref{conditionk1}) is
\begin{equation}\label{higheq}
    k_{1n}\frac{b-a}{2}\approx \frac{n\pi}{2}-\left[\frac{\pi}{2}-\sqrt{2\left(1-\frac{k_{1n}}{\beta}\right)}\right]
    =\frac{(n-1)\pi}{2}+\sqrt{2-\frac{2}{\beta}\frac{(n-1)\pi}{b-a}}
\end{equation}
where the Taylor expansion
\begin{equation}
    \arcsin(1-x)\approx \frac{\pi}{2}-\sqrt{2x}-\frac{(\sqrt{2x})^3}{24}-...
\end{equation}
has been used. The error and relative error are
\begin{equation}\begin{split}
    \Delta k_{1n}&=\frac{1}{12(b-a)}\left[2-\frac{2}{\beta}\frac{(n-1)\pi}{b-a}\right]^{3/2}\\
    \delta k_{1n}&=\frac{1}{12(n-1)\pi}\left[2-\frac{2}{\beta}\frac{(n-1)\pi}{b-a}\right]^{3/2}
\end{split}\end{equation}

Now we calculate the pricing kernel of proportional double-barrier step option. a and b in (\ref{potential}) could be considered as the lower and the upper barriers of the option. Let $\tau_1$ indicates the occupation time between the lower barrier a and the upper barrier b, and $\tau_2$ is the occupation time below the lower barrier a and above the upper barrier b. The pricing kernel is
\begin{equation}\begin{split}\label{pricing}
    p_{\rm PDBS}(x,x^\prime;\tau)&=\braket{x|e^{-\tau_1H_1-\tau_2H_2}|x^\prime}\\
    &=\int_{-\infty}^{+\infty}{\rm d}x^{\prime\prime}\braket{x|e^{-\tau_1 H_1}|x^{\prime\prime}}\braket{x^{\prime\prime}|e^{-\tau_2 H_2}|x^{\prime}}\\
    &=e^{-\tau\gamma}\int_{-\infty}^{+\infty}{\rm d}x^{\prime\prime}e^{\alpha(x-x^\prime)}\sum_n e^{-\tau_1 E_{1n}-\tau_2 E_{2n}}\phi_n(x)\phi_n(x^{\prime\prime})\phi_n(x^{\prime\prime})\phi_n(x^\prime)\\
    &=e^{-\tau\gamma}\int_{-\infty}^{+\infty}{\rm d}x^{\prime\prime}e^{\alpha(x-x^\prime)}\sum_n e^{-\frac{1}{2}\tau\sigma^2k_{1n}^2}\phi_n(x)\phi_n(x^{\prime\prime})\phi_n(x^{\prime\prime})\phi_n(x^\prime)
\end{split}\end{equation}
and the proportional double-barrier call price is 
\begin{equation}
    C_{\rm PDBS}(x;\tau)=\int_{{\rm ln}K}^{+\infty}{\rm d}x^\prime p_{\rm PDBS}(x,x^\prime;\tau) (e^{x^\prime}-K)
\end{equation}
where  
\begin{equation}\begin{split}
    H_1&=e^{\alpha x}\left(-\frac{\sigma^2}{2}\frac{\partial^2}{\partial x^2}+\gamma\right)e^{-\alpha x}\\
    H_2&=e^{\alpha x}\left(-\frac{\sigma^2}{2}\frac{\partial^2}{\partial x^2}+\gamma\right)e^{-\alpha x}+V_0
\end{split}\end{equation}
 $\phi_n(x)$ is the energy eigenstate in coordinate representation, $K$ is the exercise price, and $\tau=\tau_1+\tau_2$ is the expiration time.  
 
 Considering different wave functions (\ref{oddeq}) and (\ref{eveneq}), and different energy level formulas (\ref{loweq}) and (\ref{higheq}), $k_{1n}$ would be  divided into four cases
	\begin{equation}\begin{split}\label{lowE}
	    k_{111}&=\frac{2m\beta\pi}{\beta(b-a)+2},\ \ {\rm for}\  E\ll V_0,\ C_1(x)\\
	    k_{112}&=\frac{(2m-1)\beta\pi}{\beta(b-a)+2},\ \ {\rm for}\  E\ll V_0,\ C_2(x)
	    	\end{split}\end{equation}
	\begin{equation}\begin{split}\label{highE}	    k_{121}&=\frac{2}{b-a}\left[\frac{(2m-1)\pi}{2}+\sqrt{2-\frac{2(2m-1)\pi}{\beta(b-a)}}\right],\ \ {\rm for}\  E\approx V_0,\ C_1(x)\\
	    k_{122}&=\frac{2}{b-a}\left[(m-1)\pi+\sqrt{2-\frac{4(m-1)\pi}{\beta(b-a)}}\right],\ \ {\rm for}\  E\approx V_0,\ C_2(x)
	\end{split}\end{equation}
where $m=1,2,3,...$. The barriers $a$ and $b$ divide the integral interval into three parts: $(-\infty, a), (a,b), (b,+\infty)$. Set $x\in(a,b)$, ${\rm ln}K\in(a,b)$, and the option price expression is calculated as
\begin{equation}\begin{split}
    C_1(x;\tau)&=e^{-\tau\gamma}\int_{{\rm ln}K}^{b}{\rm d}x^\prime e^{\alpha(x-x^\prime)}\int_{-\infty}^a{\rm d}x^{\prime\prime} \bigg[\sum_{m=1}^{m_1}e^{-\frac{1}{2}\tau\sigma^2k_{111}^2}\phi_{111}(x)\phi_{211}^2(x^{\prime\prime})\bigg|_{x^{\prime\prime}<a}\phi_{111}(x^\prime)\bigg|_{{\rm ln}K<x^\prime<b}\\
    &+\sum_{m=m_1+1}^{m_{max1}}e^{-\frac{1}{2}\tau\sigma^2k_{121}^2}\phi_{121}(x)\phi_{221}^2(x^{\prime\prime})\bigg|_{x^{\prime\prime}<a}\phi_{121}(x^\prime)\bigg|_{{\rm ln}K<x^\prime<b}+\\
    &+\sum_{m=1}^{m_2}e^{-\frac{1}{2}\tau\sigma^2k_{112}^2}\phi_{112}(x)\phi_{212}^2(x^{\prime\prime})\bigg|_{x^{\prime\prime}<a}\phi_{121}(x^\prime)\bigg|_{{\rm ln}K<x^\prime<b}+\\
    &+\sum_{m=m_2+1}^{m_{max2}}e^{-\frac{1}{2}\tau\sigma^2k_{122}^2}\phi_{122}(x)\phi_{222}^2(x^{\prime\prime})\bigg|_{x^{\prime\prime}<a}\phi_{122}(x^\prime)\bigg|_{{\rm ln}K<x^\prime<b}\bigg](e^{x^\prime}-K)
    \end{split}\end{equation}
\begin{equation}\begin{split}  
    C_2(x;\tau)&=e^{\tau\gamma}\int_{b}^{+\infty}{\rm d}x^\prime e^{\alpha(x-x^\prime)}\int_{-\infty}^a{\rm d}x^{\prime\prime}\bigg[\sum_{m=1}^{m_1}e^{-\frac{1}{2}\tau\sigma^2k_{111}^2}\phi_{111}(x)\phi_{211}^2(x^{\prime\prime})\bigg|_{x^{\prime\prime}<a}\phi_{211}(x^\prime)\bigg|_{x^\prime>b}\\
    &+\sum_{m=m_2+1}^{m_{max1}}e^{-\frac{1}{2}\tau\sigma^2k_{121}^2}\phi_{121}(x)\phi_{221}^2(x^{\prime\prime})\bigg|_{x^{\prime\prime}<a}\phi_{221}(x^\prime)\bigg|_{x^\prime>b}\\
    &+\sum_{m=1}^{m_2}e^{-\frac{1}{2}\tau\sigma^2k_{112}^2}\phi_{112}(x)\phi_{212}^2(x^{\prime\prime})\bigg|_{x^{\prime\prime<a}}\phi_{212}(x^\prime)\bigg|_{x^\prime>b}\\
    &+\sum_{m=m_2+1}^{m_{max2}}e^{-\frac{1}{2}\tau\sigma^2k_{122}^2}\phi_{122}(x)\phi_{222}^2(x^{\prime\prime})\bigg|_{x^{\prime\prime}<a}\phi_{222}(x^\prime)\bigg|_{x^\prime>b}\bigg](e^{x^\prime}-K)
   \end{split}\end{equation}
\begin{equation}\begin{split}
    C_3(x;\tau)&=e^{\tau\gamma}\int_{{\rm ln}K}^b{\rm d}x^\prime e^{\alpha(x-x^\prime)}\int_a^b{\rm d}x^{\prime\prime}\bigg[\sum_{m=1}^{m_1}e^{-\frac{1}{2}\tau\sigma^2k_{111}^2}\phi_{111}(x)\phi_{111}^2(x^{\prime\prime})\bigg|_{a<x^{\prime\prime}<b}\phi_{111}(x^\prime)\bigg|_{{\rm ln}K<x^\prime<b}\\
    &+\sum_{m=m_1+1}^{m_{max1}}e^{-\frac{1}{2}\tau\sigma^2k_{121}^2}\phi_{121}(x)\phi_{121}^2(x^{\prime\prime})\bigg|_{a<x^{\prime\prime}<b}\phi_{121}(x^\prime)\bigg|_{{\rm ln}K<x^\prime<b}\\
    &+\sum_{m=1}^{m_2}e^{-\frac{1}{2}\tau\sigma^2k_{112}^2}\phi_{112}(x)\phi_{112}^2(x^{\prime\prime})\bigg|_{a<x^{\prime\prime}<b}\phi_{112}(x^\prime)\bigg|_{{\rm ln}K<x^\prime<b}\\
    &+\sum_{m_{m_2+1}}^{m_{max2}}e^{-\frac{1}{2}\tau\sigma^2k_{122}^2}\phi_{122}(x)\phi_{122}^2\bigg](x^{\prime\prime})\bigg|_{a<x^{\prime\prime}<b}\phi_{122}(x^\prime)\bigg|_{{\rm ln}K<x^\prime<b}\bigg](e^{x^\prime}-K)
   \end{split}\end{equation}
 \begin{equation}\begin{split}  
    C_4(x;\tau)&=e^{\tau\gamma}\int_{b}^{+\infty}{\rm d}x^\prime e^{\alpha(x-x^\prime)}\int_a^b{\rm d}x^{\prime\prime}\bigg[\sum_{m=1}^{m_1}e^{-\frac{1}{2}\tau\sigma^2k_{111}^2}\phi_{111}(x)\phi_{111}^2(x^{\prime\prime})\bigg|_{a<x^{\prime\prime}<b}\phi_{221}(x^\prime)\bigg|_{x^\prime>b}\\
    &+\sum_{m=m_1+1}^{m_{max1}}e^{-\frac{1}{2}\tau\sigma^2k_{121}^2}\phi_{121}(x)\phi_{121}^2(x^{\prime\prime})\bigg|_{a<x^{\prime\prime}<b}\phi_{221}(x^\prime)\bigg|_{x^\prime>b}\\
    &+\sum_{m=1}^{m_2}e^{-\frac{1}{2}\tau\sigma^2k_{112}^2}\phi_{112}(x)\phi_{112}^2(x^{\prime\prime})\bigg|_{a<x^{\prime\prime}<b}\phi_{212}(x^\prime)\bigg|_{x^\prime>b}\\
    &+\sum_{m=m_2+1}^{m_{max2}}e^{-\frac{1}{2}\tau\sigma^2k_{122}^2}\phi_{122}(x)\phi_{122}^2(x^{\prime\prime})\bigg|_{a<x^{\prime\prime}<b}\phi_{222}(x^\prime)\bigg|_{x^\prime>b}(e^{x^\prime}-K)
    \end{split}\end{equation}
\begin{equation}\begin{split}
    C_5(x;\tau)&=e^{\tau\gamma}\int_{{\rm ln}K}^b{\rm d}x^\prime e^{\alpha(x-x^\prime)}\int_b^{+\infty}{\rm d}x^{\prime\prime}\bigg[\sum_{m=1}^{m_1}e^{-\frac{1}{2}\tau\sigma^2k_{111}^2}\phi_{111}(x)\phi_{211}^2(x^{\prime\prime})\bigg|_{x^{\prime\prime}>b}\phi_{111}(x^\prime)\bigg|_{{\rm ln}K<x^\prime<b}\\
    &+\sum_{m=m_1+1}^{m_{max1}}e^{-\frac{1}{2}\tau\sigma^2k_{121}^2}\phi_{121}(x)\phi_{221}^2(x^{\prime\prime})\bigg|_{x^{\prime\prime}>b}\phi_{121}(x^\prime)\bigg|_{{\rm ln}K<x^\prime<b}\\
    &+\sum_{m=1}^{m_2}e^{-\frac{1}{2}\tau\sigma^2k_{112}^2}\phi_{112}(x)\phi_{212}^2(x^{\prime\prime})\bigg|_{x^{\prime\prime}>b}\phi_{112}(x^\prime)\bigg|_{{\rm ln}K<x^\prime<b}\\
    &+\sum_{m=m_2+1}^{m_{max2}}e^{-\frac{1}{2}\tau\sigma^2k_{122}^2}\phi_{122}(x)\phi_{222}^2(x^{\prime\prime})\bigg|_{x^{\prime\prime}>b}\phi_{122}(x^\prime)\bigg|_{{\rm ln}K<x^\prime<b}\bigg](e^{x^\prime}-K)
 \end{split}\end{equation}
 \begin{equation}\begin{split}
    C_6(x;\tau)&=e^{\tau\gamma}\int_b^{+\infty}{\rm d}x^\prime e^{\alpha(x-x^\prime)}\int_b^{+\infty}{\rm d}x^{\prime\prime}\bigg[\sum_{m=1}^{m_1}e^{-\frac{1}{2}\tau\sigma^2k_{111}^2}\phi_{111}(x)\phi_{211}^2(x^{\prime\prime})\bigg|_{x^{\prime\prime}>b}\phi_{211}(x^\prime)\bigg|_{x^\prime>b}\\
    &+\sum_{m=m_1+1}^{m_{max1}}e^{-\frac{1}{2}\tau\sigma^2k_{121}^2}\phi_{121}(x)\phi_{221}^2(x^{\prime\prime})\bigg|_{x^{\prime\prime}>b}\phi_{221}(x^\prime)\bigg|_{x^\prime>b}\\
    &+\sum_{m=1}^{m_2}e^{-\frac{1}{2}\tau\sigma^2k_{112}^2}\phi_{112}(x)\phi_{212}^2(x^{\prime\prime})\bigg|_{x^{\prime\prime}>b}\phi_{212}(x^\prime)\bigg|_{x^\prime>b}\\
    &+\sum_{m=m_2+1}^{m_{max2}}e^{-\frac{1}{2}\tau\sigma^2k_{122}^2}\phi_{122}(x)\phi_{222}^2(x^{\prime\prime})\bigg|_{x^{\prime\prime}>b}\phi_{222}(x^\prime)\bigg|_{x^\prime>b}\bigg](e^{x^\prime}-K)
 \end{split}\end{equation}
 the option price is
 \begin{equation}
       C_{\rm PDBS}(x;\tau)=C_1(x;\tau)+C_2(x;\tau)+C_3(x;\tau)+C_4(x;\tau)+C_5(x;\tau)+C_6(x;\tau)
 \end{equation}
 where
 \begin{equation}\begin{split}
   \phi_{111}(x)&=A_{111}\sin\bigg[k_{111}\bigg(x-\frac{b+a}{2}\bigg)\bigg]\\
   \phi_{121}(x)&=A_{121}\sin\bigg[k_{121}\bigg(x-\frac{b+a}{2}\bigg)\bigg]\\
   \phi_{112}(x)&=A_{112}\cos\bigg[k_{112}\bigg(x-\frac{b+a}{2}\bigg)\bigg]\\
   \phi_{122}(x)&=A_{122}\cos\bigg[k_{122}\bigg(x-\frac{b+a}{2}\bigg)\bigg]
    \end{split}\end{equation}
 \begin{equation}\begin{split}  
   \phi_{211}(x)\bigg|_{x>a}&=A_{211}e^{-k_{211}\left(x-\frac{b+a}{2}\right)},\ \ \phi_{211}(x)\bigg|_{x<a}=-A_{211}e^{k_{211}\left(x-\frac{b+a}{2}\right)}\\
   \phi_{221}(x)\bigg|_{x>a}&=A_{221}e^{-k_{221}\left(x-\frac{b+a}{2}\right)},\ \ 
   \phi_{221}(x)\bigg|_{x<a}=-A_{221}e^{k_{221}\left(x-\frac{b+a}{2}\right)}\\
   \phi_{212}(x)\bigg|_{x>a}&=A_{212}e^{-k_{212}\left(x-\frac{b+a}{2}\right)},\ \
   \phi_{212}(x)\bigg|_{x<a}=A_{212}e^{k_{212}\left(x-\frac{b+a}{2}\right)}\\
   \phi_{222}(x)\bigg|_{x>a}&=A_{222}e^{-k_{222}\left(x-\frac{b+a}{2}\right)},\ \
   \phi_{222}(x)\bigg|_{x<a}=A_{222}e^{k_{222}\left(x-\frac{b+a}{2}\right)}
 \end{split}\end{equation}
 \begin{equation}\begin{split}\label{k2}
 k_{211}&=\sqrt{\frac{2V_0}{\sigma^2}-k_{111}^2},\ \ k_{221}=\sqrt{\frac{2V_0}{\sigma^2}-k_{121}^2}\\
 k_{212}&=\sqrt{\frac{2V_0}{\sigma^2}-k_{112}^2},\ \ k_{222}=\sqrt{\frac{2V_0}{\sigma^2}-k_{122}^2}
  \end{split}\end{equation}
  \begin{equation}\begin{split}\label{A1}
      A_{111}&=\sqrt{\frac{2k_{211}}{k_{211}(b-a)+2}},\ \ A_{112}=\sqrt{\frac{2k_{212}}{k_{212}(b-a)+2}}\\
      A_{121}&=\sqrt{\frac{2k_{221}}{k_{221}(b-a)+2}},\ \ 
      A_{122}=\sqrt{\frac{2k_{222}}{k_{222}(b-a)+2}}
  \end{split}\end{equation}
 \begin{equation}\begin{split}\label{A2}
 A_{211}&=A_{111}\sin\bigg(k_{111}\frac{b-a}{2}\bigg)e^{k_{211}\frac{b-a}{2}},\ \
 A_{212}=A_{112}\cos\bigg(k_{112}\frac{b-a}{2}\bigg)e^{k_{212}\frac{b-a}{2}}\\
 A_{221}&=A_{121}\sin\bigg(k_{121}\frac{b-a}{2}\bigg)e^{k_{221}\frac{b-a}{2}},\ \ 
 A_{222}=A_{122}\cos\bigg(k_{122}\frac{b-a}{2}\bigg)e^{k_{222}\frac{b-a}{2}}
  \end{split}\end{equation}
(\ref{k2}),(\ref{A1}) and (\ref{A2}) are derived from (\ref{A1A2}) and (\ref{beta}). $m_{max1}$ and $m_{max2}$ are corresponding to the maximum value of $m$ in high energy $(E\approx V_0)$\ formulas (\ref{highE}), $m_1$ and $m_2$ are corresponding to the maximum value of $m$ in low energy $(E\ll V_0)$\ formulas (\ref{lowE}), respectively. The option price for $x<a$ and $x>b$ could be obtained similarly.

\begin{table}

		\begin{center}
			\begin{tabular}{|c|c|c|}
				\hline
				energy level $n$  &    relative error for low energy formula(\ref{lowE}) & relative error for high energy formula(\ref{highE})  \\
				\hline

				$n=1$    &   $2.14\times 10^{-4}$  & 0.0833 \\
				\hline
				
		  	    $n=2$     &   $8.55\times 10^{-4}$&  0.0276 \\
				\hline
				
			 $n=3$    &   0.002&  0.01 \\
				\hline
				
                $n=4$     &   0.0034&  0.00296\\
				\hline
			\end{tabular}
		\end{center}
		\caption{relative errors for  different energy levels at $V_0=55$. Parameters: $a={\rm ln}90=4.5$, $b={\rm ln}130=4.867$, $V_0=55$, $\sigma=0.3$, $r=0.05$.}
		\label{table: table1}
	\end{table}

In Table.~\ref{table: table1}, the relative errors for (\ref{lowE}) and (\ref{highE}) are shown. For $a={\rm ln}90=4.5$, $b={\rm ln}130=4.867$,$V_0=55$, $\sigma=0.3$ and $r=0.05$, we have $\beta=34.96$ and $n_{max}=4$. It is shown that, with the increasing of $n$, the error in the second column increases while the error in the third column decreases. For low energy levels $(n\leq 3)$, the 
error of (\ref{lowE}) is smaller, and for high energy level $(n= 4)$, the error of (\ref{highE}) is smaller.   
\begin{table}

		\begin{center}
			\begin{tabular}{|p{2cm}|p{2cm}|p{2cm}|p{2cm}|p{2cm}|p{2cm}|p{2cm}|}
				\hline
				\makecell[c]{$V_0$}  &    \makecell[c]{$\beta$} & \makecell[c]{$n_{max}$}  & \makecell[c]{$m_{max1}$} &  \makecell[c]{$m_{max2}$}  & \makecell[c]{$m_1$}  & \makecell[c]{$m_2$}\\
				\hline

				\makecell[c]{55}    &   \makecell[c]{35}  & \makecell[c]{4} &\makecell[c]{2} &\makecell[c]{-} &\makecell[c]{1} &\makecell[c]{2} \\
				\hline
				
		  	   \makecell[c]{26}    &   \makecell[c]{24}  & \makecell[c]{2} &\makecell[c]{1} &\makecell[c]{-} &\makecell[c]{-} &\makecell[c]{1} \\
				\hline
				
			 	\makecell[c]{13}    &  \makecell[c]{17}  & \makecell[c]{1} &\makecell[c]{-} &\makecell[c]{-} &\makecell[c]{-} &\makecell[c]{1} \\
				\hline
			\end{tabular}
		\end{center}
		\caption{$m_{max1}$, $m_{max2}$ and $m_1$, $m_2$ for $V_0=55, 26, 12$, respectively. Parameters: $a={\rm ln}90=4.5$, $b={\rm ln}130=4.867$.}
			\label{table: table2}
	\end{table}
In Table.~\ref{table: table2}, $m_{max1}$, $m_{max2}$ and $m_1$, $m_2$ for $V_0=55, 26, 13$ (or daily knock-out factors $d=0.8, 0.9, 0.95$~\cite{Linetsky:2001}) are shown, respectively.We will use different energy formulas for different $m$ to give a more accurate pricing kernel.
\section{Numerical Results}
\begin{figure}[!htbp]
\begin{center}
\includegraphics[width=0.49\linewidth]{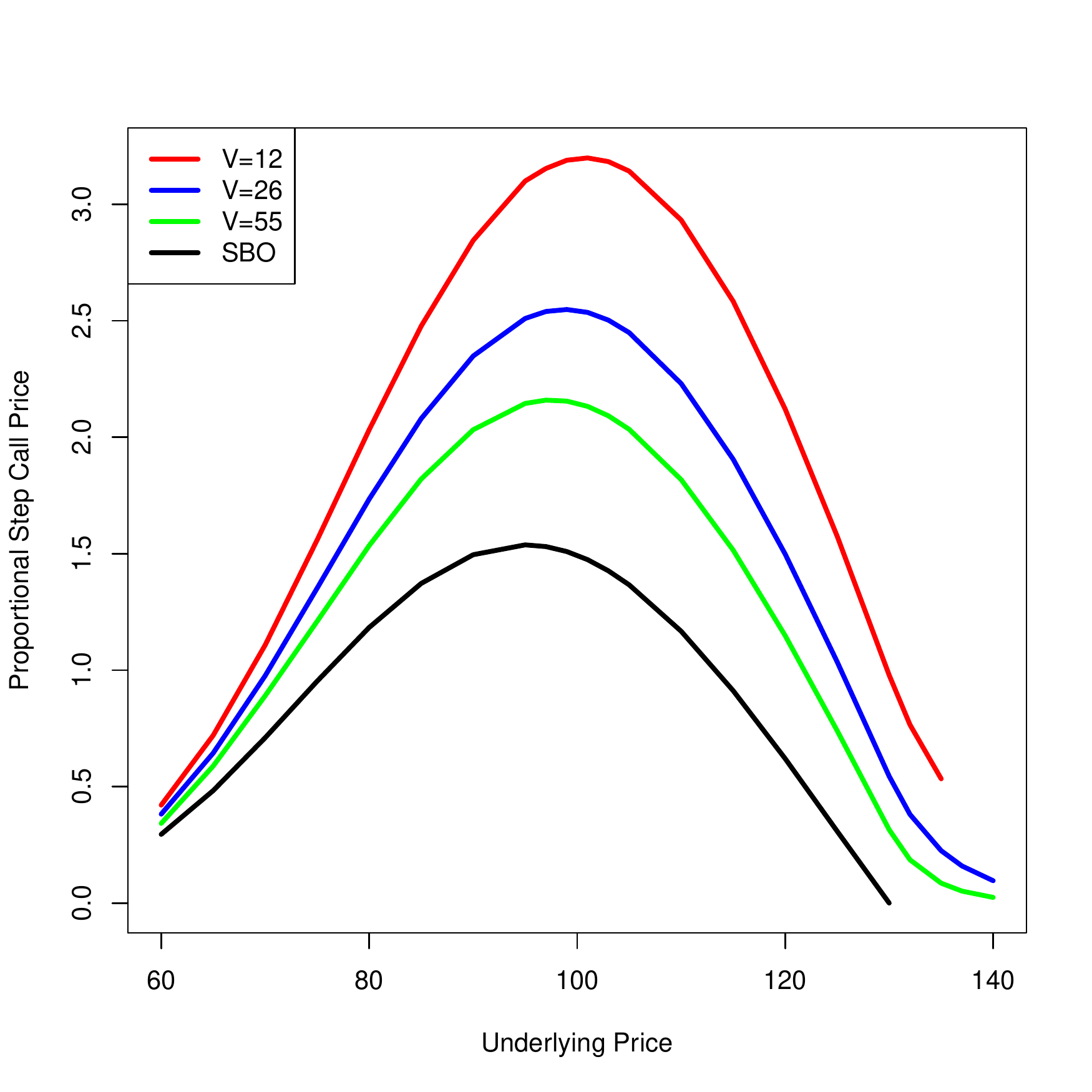}
\includegraphics[width=0.49\linewidth]{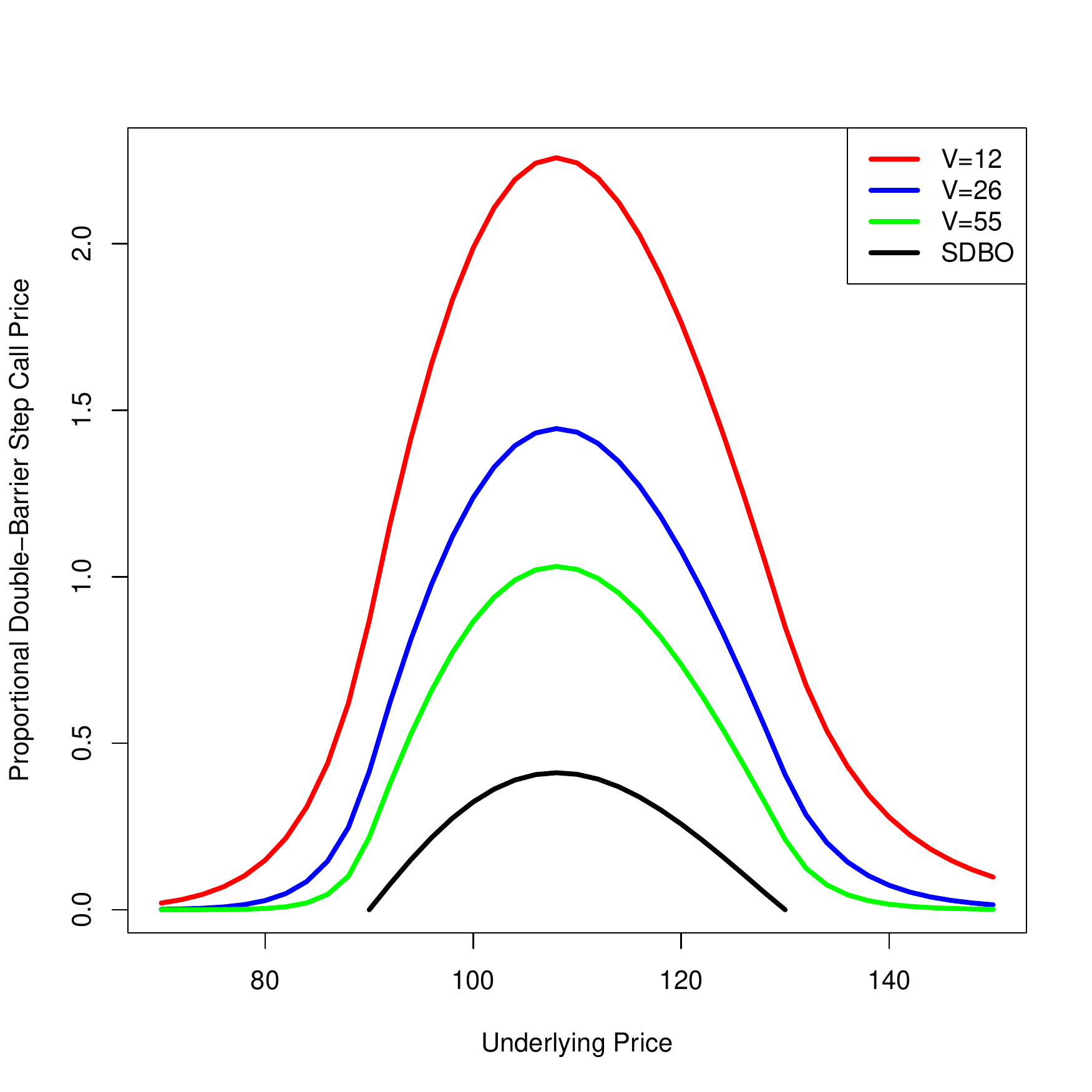}
\end{center}
\caption{Proportional step call price (left) and proportional double-barrier step call price (right) as functions of underlying price for different potentials, respectively. Parameters: $a=4.5$, $B=b=4.867$, $K=100$, $r=0.05$, $\sigma=0.3$, $\tau=1$.}
\label{fig: price-underlying}
\end{figure}
In Fig.~\ref{fig: price-underlying}, we show the proportional step call price and the proportional double-barrier step call price as functions of underlying price, respectively. The black lines are corresponding to the standard step option (SBO) and the standard double-barrier option (SDBO) for comparison. It is shown that the option prices decrease with the increasing of potential $V_0$ for both the two diagrams. In the limit $V_0\to\infty$, the option payoff tends to be the payoff of a standard step option (left) or of a standard double-barrier step option. The results agree with the results given in~\cite{Linetsky:1999,Linetsky:2001}.
\begin{figure}[!htbp]
\begin{center}
\includegraphics[width=0.49\linewidth]{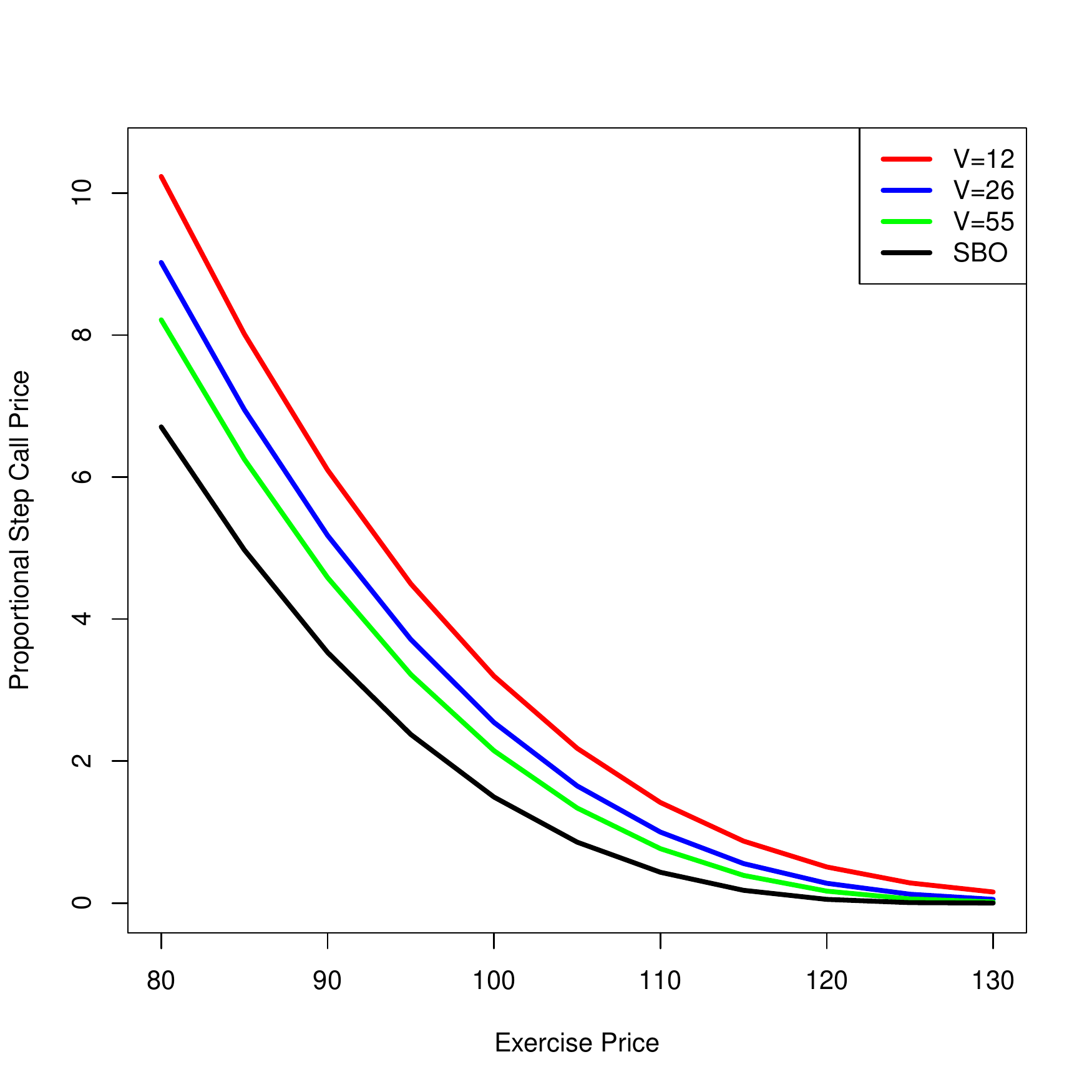}
\includegraphics[width=0.49\linewidth]{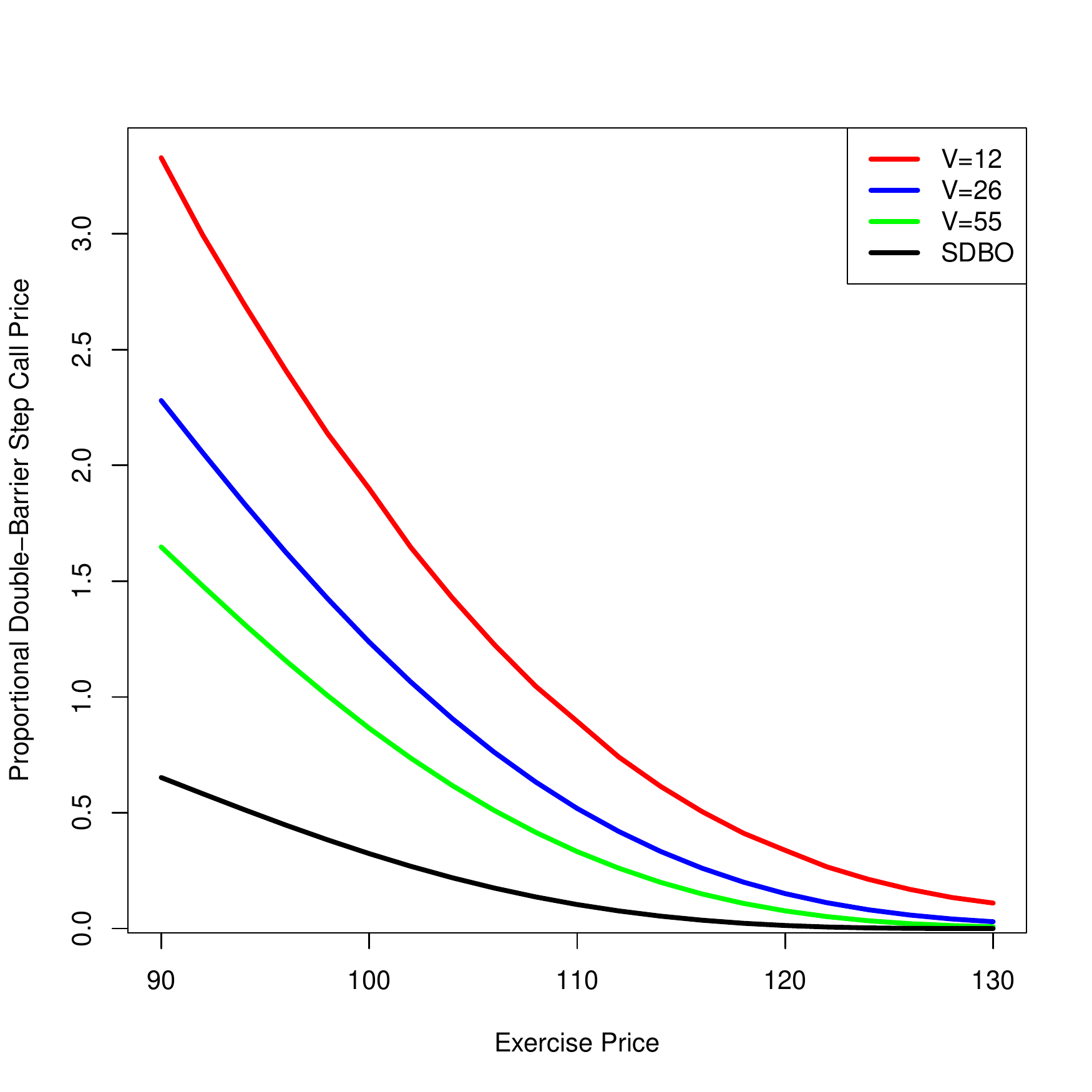}
\end{center}
\caption{Proportional step call price (left) and proportional double-barrier step call price (right) as functions of  exercise price for different  potentials, respectively. Parameters: $a=4.5$, $B=b=4.867$, $x=4.605$, $r=0.05$, $\sigma=0.3$, $\tau=1$.}
\label{fig: price-exercise}
\end{figure}

In Fig.~\ref{fig: price-exercise}, we show the proportional step call price and the proportional double-barrier step call price as functions of exercise price. The black lines are corresponding to the standard step option (SBO) and the standard double-barrier option (SDBO) for comparison. In the limit $V_0\to\infty$, the option payoff tends to be the payoff of a standard barrier option or a standard double-barrier option. For a fixed $V_0$, the option price decreases with the increasing of exercise price $K$.
\begin{figure}[!htbp]
\begin{center}
\includegraphics[width=0.49\linewidth]{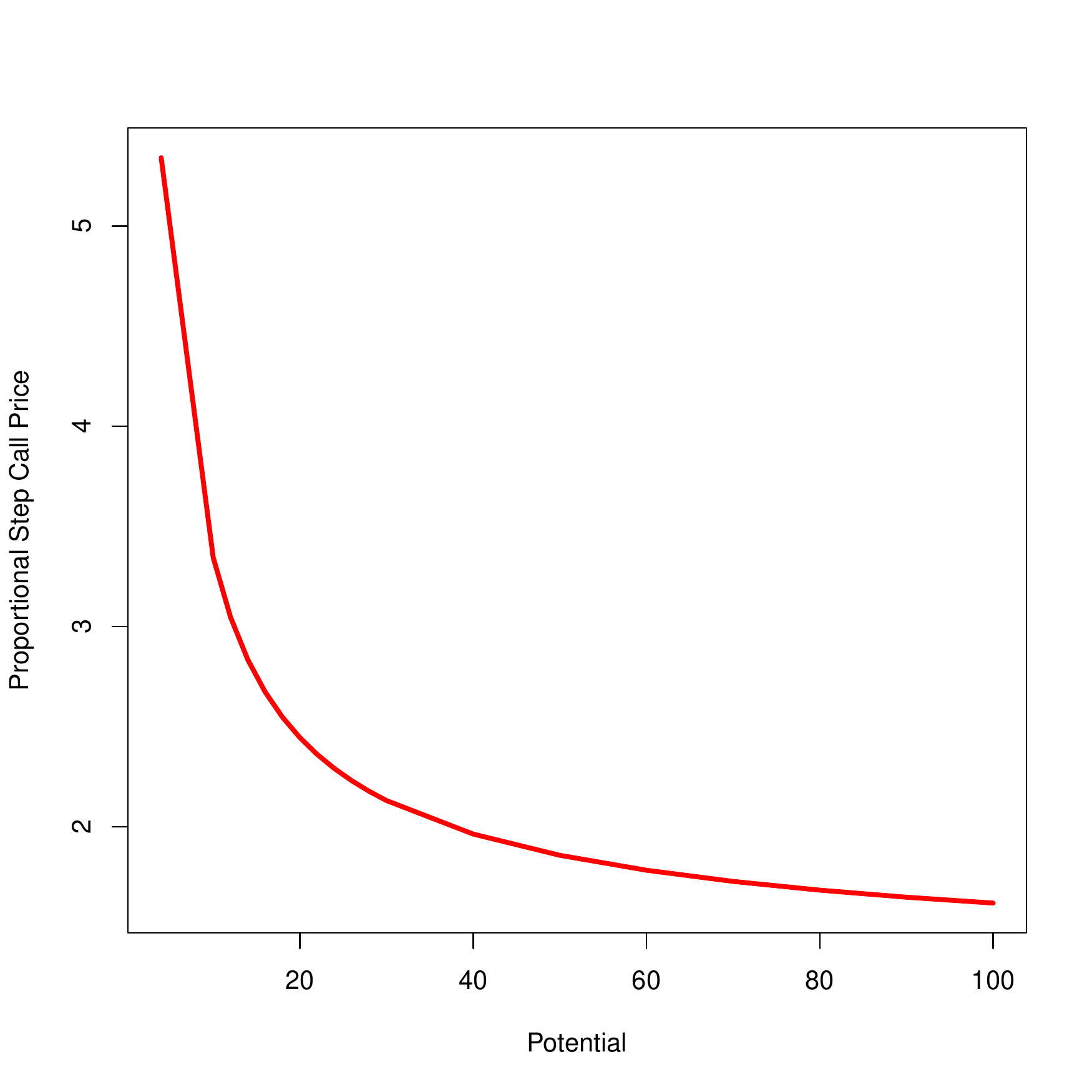}
\includegraphics[width=0.49\linewidth]{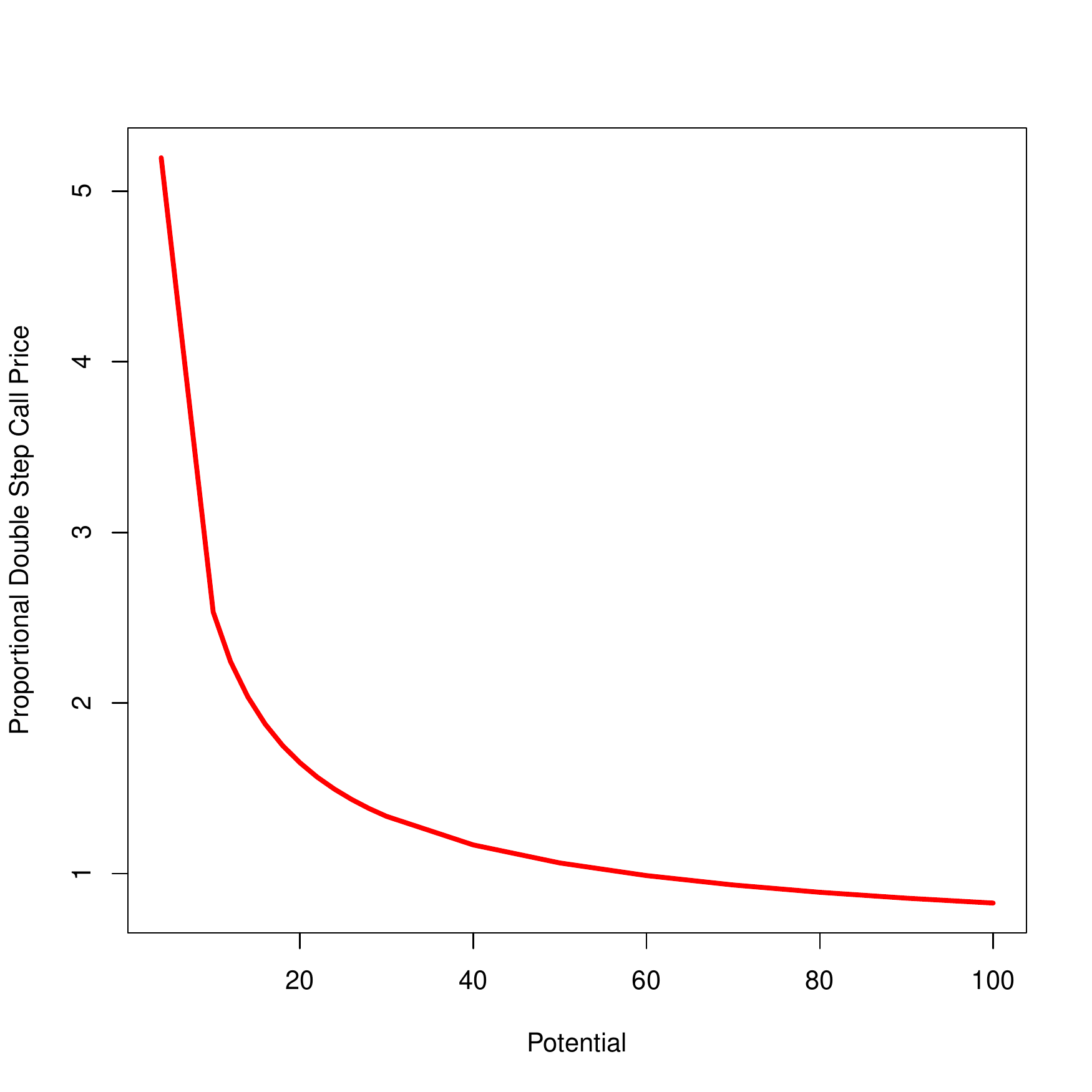}
\end{center}
\caption{Proportional step call price (left) and proportional double-barrier step call price (right) as functions of potential . Parameters: $a=4.5$, $B=b=4.867$, $S=110$, $K=100$, $r=0.05$, $\sigma=0.3$, $\tau=1$.}
\label{fig: price-potential}
\end{figure}

In Fig.~\ref{fig: price-potential}, the proportional step call price and the proportional double-barrier step call price as functions of potential $V_0$ are shown. The results are  consistent with the results given in~\cite{Linetsky:1999,Linetsky:2001}, where the independent variable is the daily knock-out factor $d=e^{-V_0/250}$. 

\begin{figure}[!htbp]
\begin{center}
\includegraphics[width=0.49\linewidth]{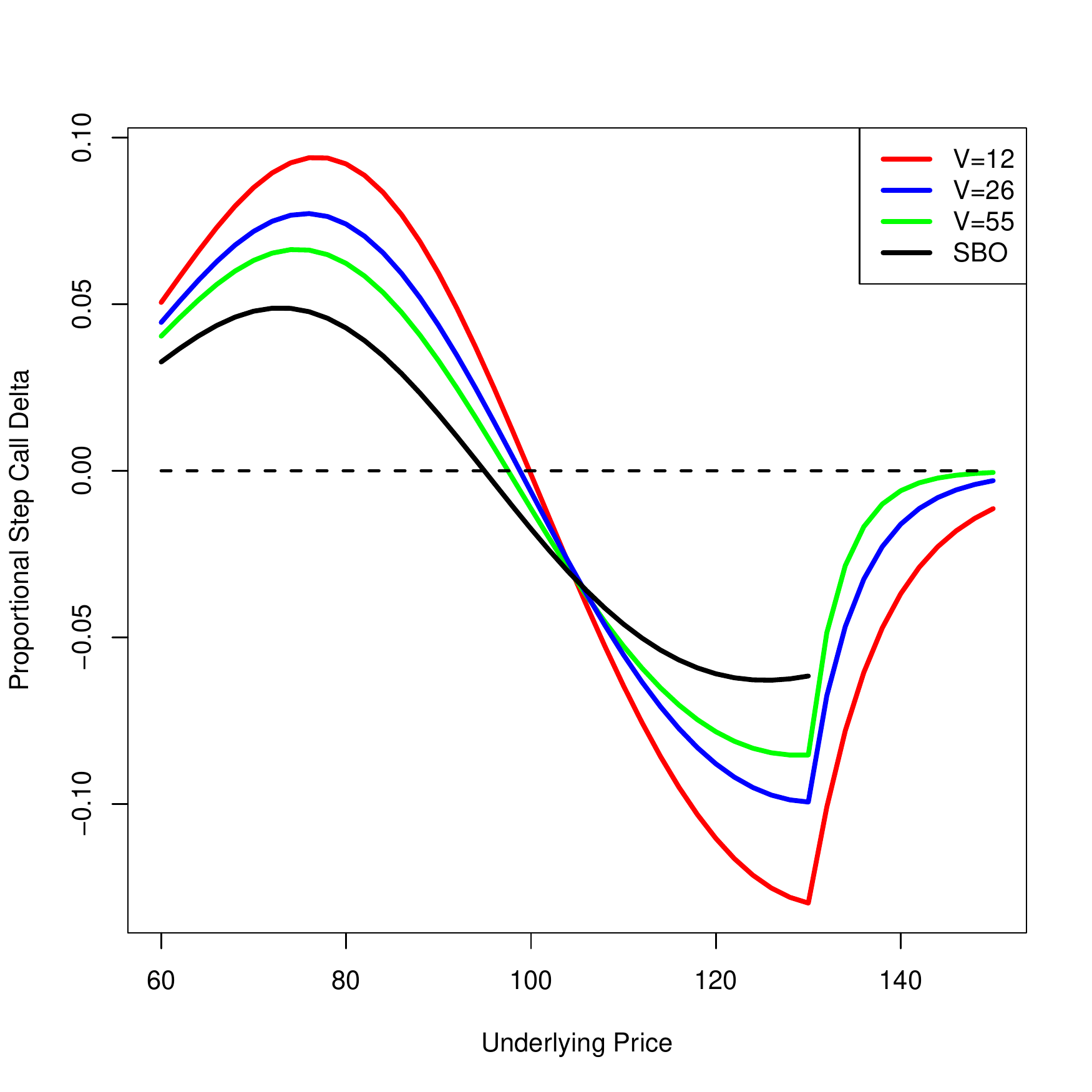}
\includegraphics[width=0.49\linewidth]{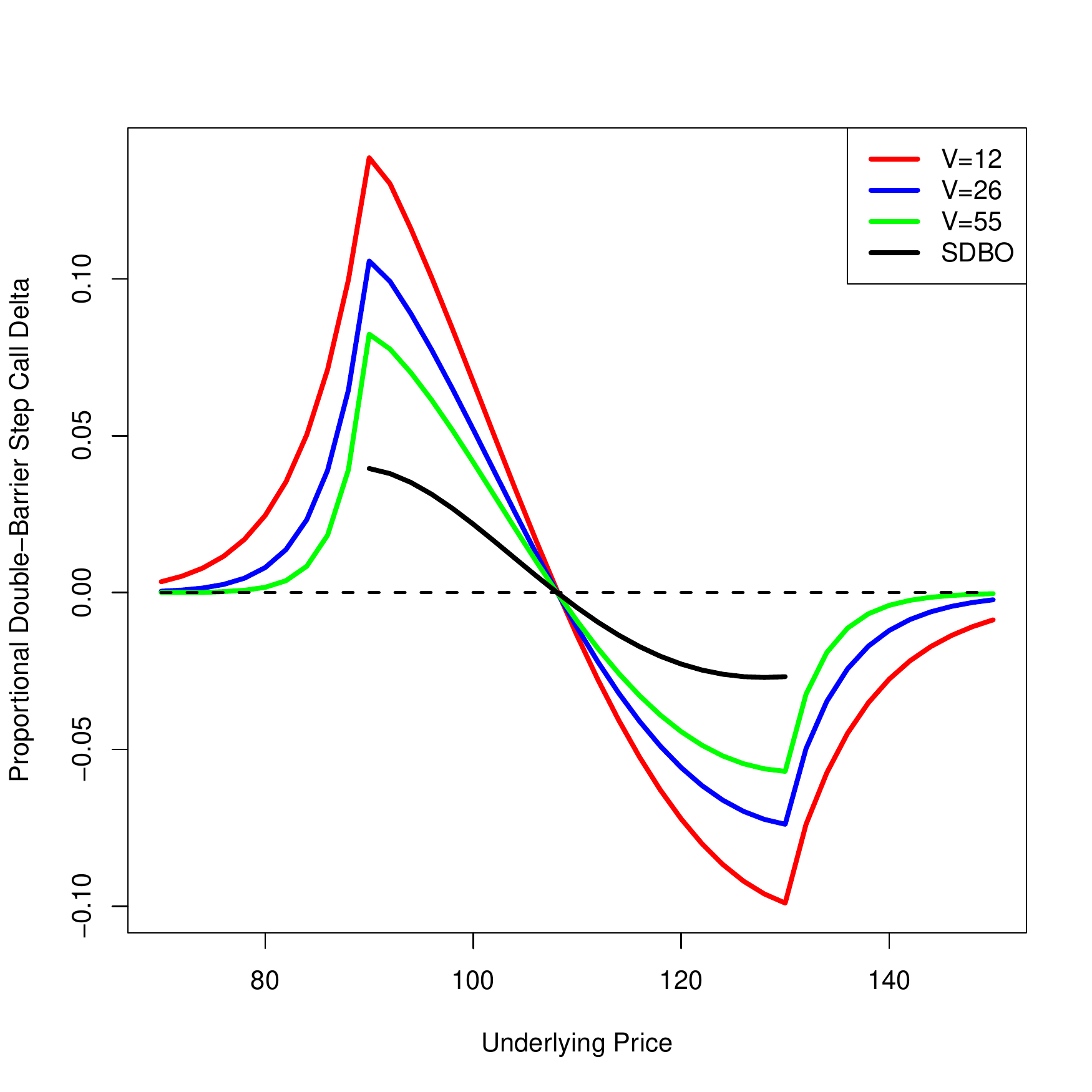}
\end{center}
\caption{Proportional step call delta (left) and proportional double-barrier step call delta (right) against underlying price for different potentials. Parameters: $a=4.5$, $B=b=4.867$, $K=100$, $x=4.605$, $r=0.05$, $\sigma=0.3$, $\tau=1$.}
\label{fig: delta-price}
\end{figure}

In Fig.~\ref{fig: delta-price}, we show the proportional step call delta and the proportional double step call delta as functions of the initial underlying price. The definition of delta is
\begin{equation}
    \Delta=\frac{\partial C}{\partial S}=e^{-x}\frac{\partial C}{\partial x}
\end{equation}
and our results are consistent with the results in~\cite{Linetsky:1999,Linetsky:2001}. The black lines in Fig.~\ref{fig: delta-price} are corresponding to the standard barrier call delta and standard double-barrier call delta for comparison, respectively.  
\section{Conclusion}

Path-integral is an effective method linking option price changing to a particle moving under some potential in the space. Here we have studied pricing of the proportional step call option and the proportional double-barrier step call option, which could be analogous to a particle moving through a trapezoid potential barrier or a symmetric square potential well, respectively. We have presented option prices changing with the initial underlying prices, different potentials, and exercise prices, respectively. The numerical results are in accordance with the results using mathematical method in~\cite{Linetsky:2001,Linetsky:1999}. 
  The pricing of other barrier options could be studied by defining appropriate potentials $V$.

\appendix
\section{Path Integral Method for Black-Scholes Model Pricing}

According to Ref~\cite{Baaquie:2004}, starting from Black-Scholes pricing formula, the price of European option can be derived by path integral method. The Black-Scholes formula is
\begin{equation}\label{eq:BS}
    \frac{\partial C}{\partial t}+rS\frac{\partial C}{\partial S}+\frac 12\sigma^2 S^2\frac{\partial^2 C}{\partial S^2}=0
\end{equation}
where $C$ is European option price, $S$ is the underlying asset price, $\sigma$ is the fixed volatility, and $r$ is the interest rate. Let
\begin{equation}
    S=e^x,\ \ (-\infty<x<+\infty)
\end{equation}
and (\ref{eq:BS}) can be denoted as
\begin{equation}
  \frac{\partial C}{\partial t}=\bigg[-\frac{\sigma^2}{2}\frac{\partial^2}{\partial x^2}+\left(\frac 12\sigma^2-r\right)+r\bigg]C  
\end{equation}
let
\begin{equation}\label{eq:BSHami}
    H_{\rm BS}=-\frac{\sigma^2}{2}\frac{\partial^2}{\partial x^2}+\left(\frac 12\sigma^2-r\right)\frac{\partial}{\partial x}+r
\end{equation}
the Black-Scholes equation is written as 
\begin{equation}\label{eq:Schro}
    \frac{\partial C}{\partial t}=H_{\rm BS}C
\end{equation}

Comparing (\ref{eq:Schro}) to Schr{\"o}dinger equation, we have
\begin{equation}
    \sigma^2\sim\frac{1}{m^2},\ \ C\sim \psi(x)
\end{equation}
where $m$ is the particle mass, and $\psi(x)$ is the wave function. The Black-Scholes Hamitonian (\ref{eq:BSHami}) in momentum representation can be denoted as
\begin{equation}
    H_{\rm BS}=\frac 12\sigma^2 p^2 +i\left(\frac 12\sigma^2-r \right)p+r
\end{equation}\label{eq:momentum}
where $p=-i\frac{\partial}{\partial x}$. The pricing kernel is
\begin{equation}\begin{split}
  \braket{x|e^{-\tau H_{\rm BS}}|x^\prime}&=\int_{-\infty}^{+\infty} \frac{{\rm d}p}{2\pi} \braket{x|e^{-\tau H_{\rm BS}}|p} \braket{p|x^\prime}\\
  &=e^{-r\tau}\int_{-\infty}^{+\infty}\frac{{\rm d}p}{2\pi}e^{-\frac 12\tau\sigma^2\left(p-\frac{x^\prime-x_0}{\tau\sigma^2}\right)^2-\frac{(x^\prime-x_0)^2}{2\tau\sigma^2}}\\
  &=\frac{1}{\sqrt{2\pi\tau\sigma^2}}e^{-r\tau}e^{-\frac{1}{2\tau\sigma^2}(x^\prime-x_0)^2}
\end{split}\end{equation}
where the completeness relation has been used, and
\begin{equation}
    x_0=x+\tau\left(r-\frac{\sigma^2}{2}\right)
\end{equation}

The European call option price can be denoted as
\begin{equation}\begin{split}
 C(x,\tau)&=e^{-r\tau}\int_{-\infty}^{+\infty}\frac{{\rm d}x^\prime}{\sqrt{2\pi\tau\sigma^2}}(e^{x^\prime}-K)_{+}e^{-\frac{1}{2\tau\sigma^2}(x^\prime-x_0)^2}\\ 
 &=e^{-r\tau}\int_{\ln{K}-x_0}^{+\infty}\frac{{\rm d}x^\prime}{\sqrt{2\pi\tau\sigma^2}}(e^{x^\prime+x_0}-K)e^{-\frac{1}{2\tau\sigma^2}{x^\prime}^2}\\
 &=SN(d_+)-e^{-r\tau}KN(d_-)
\end{split}\end{equation}
where
\begin{equation}
    N(x)=\frac{1}{\sqrt{2\pi}}\int_{-\infty}^{x}e^{-\frac 12 z^2}{\rm d}z,\ \ d_{\pm}=\frac{\ln{\frac{S}{K}}+\left(r\pm\frac{\sigma^2}{2}\right)\tau}{\sigma\sqrt{\tau}}
\end{equation}
\section{Path Integral Method for the Standard Barrier Option Pricing}
The up-and-out standard barrier (UOSB) option Hamiltonian is
\begin{equation}\begin{split}
H_{\rm UOSB}&=H_{\rm BS}+V(x)\\
&=-\frac{\sigma^2}{2}\frac{\partial^2}{\partial x^2}+\left(\frac{\sigma^2}{2}-r\right)\frac{\partial}{\partial x}+r+V(x)\\
&=e^{\alpha x}\left(-\frac{\sigma^2}{2}\frac{\partial^2}{\partial x^2}+\gamma\right)e^{-\alpha x}+V(x)
\end{split}\end{equation}
where 
\begin{equation}
    \alpha=\frac{1}{\sigma^2}\left(\frac{\sigma^2}{2}-r\right),\ \ \gamma=\frac{1}{2\sigma^2}\left(\frac{\sigma^2}{2}+r\right)^2
\end{equation}
and the potential $V(x)$ is
\begin{equation}
V(x)=\left\{
\begin{aligned}
0 & , & x< B,\\
\infty & , & x\geq B.
\end{aligned}
\right.
\end{equation}
the corresponding wave function is
\begin{equation}
 C(x)=\left\{
\begin{aligned}
&e^{ip(x-B)}-e^{-ip(x-B)}  , & x<B,\\
&0  , & x\geq B.
\end{aligned}
\right.   
\end{equation}
and the pricing kernel is
\begin{equation}\begin{split}
p_{\rm UOSB}(x,x^\prime;\tau)&=\braket{x|e^{-\tau H_{DB}}|x^\prime}\\
&=e^{-\tau\gamma}e^{\alpha(x-x^\prime)}\int_{0}^{\infty}\frac{{\rm d}p}{2\pi}e^{-\frac{1}{2}\tau\sigma^2p^2}\big[e^{ip(x-B)}-e^{-ip(x-B)}\big]\big[e^{-ip(x^\prime-B)-e^{ip(x^\prime-B)}}\big]\\
&=2e^{-\tau\gamma}e^{\alpha(x-x^\prime)}\int_{0}^{\infty}\frac{{\rm d}p}{2\pi}e^{-\frac{1}{2}\tau\sigma^2p^2}[\cos(p(x-x^\prime))-\cos(p(x+x^\prime-2B))]
\end{split}\end{equation}
the corresponding option price
\begin{equation}\begin{split}
    C_{\rm UOSB}(x;\tau)&=\int_{{\rm ln}K}^B {\rm d}x^\prime p_{\rm UOSB}(x,x^\prime;\tau)(e^{x^\prime}-K)\\
    &=2e^{-\tau\gamma}\int_{{\rm ln}K}^B {{\rm d}x^\prime}e^{\alpha(x-x^\prime)}\int_0^{\infty}\frac{{\rm d}p}{2\pi}e^{-\frac{1}{2}\tau\sigma^2p^2}[\cos(p(x-x^\prime))-\cos(p(x+x^\prime-2B))]
\end{split}\end{equation}
\section{Path Integral Method for the Standard Double-Barrier Option Pricing}
The standard double-barrier (SDB) option Hamiltonian is~\cite{Baaquie:2004}
\begin{equation}\begin{split}
    H_{\rm SDB}&=H_{\rm BS}+V(x)\\
    &=-\frac{\sigma^2}{2}\frac{\partial^2}{\partial x^2}+\left(\frac{\sigma^2}{2}-r\right)\frac{\partial}{\partial x}+r+V(x)\\
&=e^{\alpha x}\left(-\frac{\sigma^2}{2}\frac{\partial^2}{\partial x^2}+\gamma\right)e^{-\alpha x}+V(x)
\end{split}\end{equation}
where 
\begin{equation}
    \alpha=\frac{1}{\sigma^2}\left(\frac{\sigma^2}{2}-r\right),\ \ \gamma=\frac{1}{2\sigma^2}\left(\frac{\sigma^2}{2}+r\right)^2
\end{equation}
and the potential $V(x)$ is
\begin{equation}
V(x)=\left\{
\begin{aligned}
\infty & , & x\leq a,\\
0 & , & a<x<b,\\
\infty & , & x\geq b.
\end{aligned}
\right.
\end{equation}
the corresponding eigenstate is
\begin{equation}
 \phi_n(x)=\left\{
\begin{aligned}
\sqrt{\frac{n\pi}{b-a}}\sin&[p_n(x-a)]  , & a<x<b,\\
0 & , & x<a,\ x>b.
\end{aligned}
\right.   
\end{equation}
where 
\begin{equation}
    p_n=\frac{n\pi}{b-a},\ \ E_n=\frac{1}{2}\sigma^2p_n^2,\ \ n=1,2,3,...
\end{equation}

The pricing kernel is
\begin{equation}\begin{split}
    p_{\rm SDB}(x,x^\prime;\tau)&=\braket{x|e^{-\tau H_{DB}}|x^\prime}\\
    &=e^{\alpha(x-x^\prime)}\braket{x|e^{-\tau\left(-\frac{\sigma^2}{2}\frac{\partial^2}{\partial x^2}+\gamma+V\right)}|x^\prime}\\
    &=e^{-\tau\gamma}e^{\alpha(x-x^\prime)}\sum_{n=1}^{+\infty}e^{-\frac{1}{2}\tau\sigma^2p_n^2}\phi_n(x)\phi_n(x^\prime)
\end{split}\end{equation}
and the option price 
\begin{equation}\begin{split}
C_{\rm SDB}(x;\tau)&=\int_{{\rm ln}K}^b{\rm d}x^\prime p_{\rm SDB}(x,x^\prime;\tau)(e^{x^\prime}-K)\\
&=\frac{2}{b-a}e^{-\tau\gamma}\int_{{\rm ln}K}^b{\rm d}x^\prime e^{\alpha(x-x^\prime)}\sum_{n=1}^{+\infty}e^{-\frac{1}{2}\tau\sigma^2\frac{n^2\pi^2}{(b-a)^2}}\sin\left[\frac{n\pi}{b-a}(x-a)\right]\sin\left[\frac{n\pi}{b-a}(x^\prime-a)\right](e^{\prime}-K)
\end{split}\end{equation}

\end{document}